\documentclass[prl,floatfix,showpacs,twocolumn,preprintnumbers,amsmath,amssymb,superscriptaddress]{revtex4}
\usepackage[sort&compress]{natbib}
\usepackage{graphicx,float,datetime}
\usepackage{hyperref,wasysym}
\usepackage{epstopdf,subfigure}
\usepackage{enumitem}
\usepackage{bm}
\usepackage[utf8]{inputenc}

\newcommand*\xbar[1]{%
  \hbox{%
    \vbox{%
      \hrule height 0.5pt % The actual bar
      \kern0.5ex%         % Distance between bar and symbol
      \hbox{%
        \kern-0.1em%      % Shortening on the left side
        \ensuremath{#1}%
        \kern-0.1em%      % Shortening on the right side
      }%
    }%
  }%
} 

\newcommand{\udt}[3]{#1^{#2}_{\phantom{#2}#3}}

\newcommand{\dut}[3]{#1_{#2}^{\phantom{#2}#3}}

\def\ber{\begin{eqnarray}}
\def\eer{\end{eqnarray}}
\def\beq{\begin{equation}}
\def\eeq{\end{equation}}

\newdateformat{mydate}{\THEDAY\hspace{3pt}\monthname[\THEMONTH] \THEYEAR}

\newcommand\scalemath[2]{\scalebox{#1}{\mbox{\ensuremath{\displaystyle #2}}}}

\begin{document}

\begin{center}
\title{Solar System tests in $f(T)$ gravity}
\date{\mydate\today}
\author{Gabriel Farrugia\footnote{gabriel.farrugia.11@um.edu.mt}}
\affiliation{Department of Physics, University of Malta, Msida, MSD 2080, Malta}
\affiliation{Institute of Space Sciences and Astronomy, University of Malta, Msida, MSD 2080, Malta}
\author{Jackson Levi Said\footnote{jackson.said@um.edu.mt}}
\affiliation{Department of Physics, University of Malta, Msida, MSD 2080, Malta}
\affiliation{Institute of Space Sciences and Astronomy, University of Malta, Msida, MSD 2080, Malta}
\author{Matteo Luca Ruggiero\footnote{matteo.ruggiero@polito.it}}
\affiliation{DISAT, Politecnico di Torino, Corso Duca degli Abruzzi 24, Torino, Italy}
\affiliation{INFN, Sezione di Torino, Via Pietro Giuria 1, Torino, Italy}
\affiliation{Dipartimento di Fisica, Universit\`{a} di Torino, Via Pietro Giuria 1, 10125 Torino, Italy}

\begin{abstract}
{
We investigate the four solar system tests of gravity - perihelion precession, light bending, Shapiro time delay, gravitational redshift - in $f(T)$ gravity. In particular, we investigate the solution derived by Ruggiero and Radicella \cite{Ruggiero:2015oka} for a nondiagonal vierbein field for a polynomial $f(T) = T + \alpha T^n$, where $\alpha$ is a constant and $|n| \neq 1$. In this paper, we derive the solutions for each test, in which Weinberg's, Bodenner and Will's, Cattani \textit{et al.} and Rindler and Ishak's methods are applied, \textit{Gravitation and Cosmology: Principles and Applications of the General Theory of Relativity} (Wiley, New York, 1972); Am. J. Phys. \textbf{71} (2003); Phys. Rev. D \textbf{87}, 047503 (2013); Phys. Rev. D \textbf{76}, 043006 (2007). We set a constraint on $\alpha$ for $n = 2, 3$ by using data available from literature.
}
\end{abstract}


\maketitle

\end{center}

%------------------------Section-------------------------
\section{I. Introduction}\label{sec:intro}
%------------------------Section-------------------------

One of the current issues in theoretical physics is the difficult to understand gravitational interactions at large scales. For instance, the observed accelerated expansion of the Universe \cite{Riess:1998cb,Perlmutter:1998np,Hinshaw:2012aka,Eisenstein:2005su,Wang:2008vja}  cannot be explained in general relativity (GR)  unless the existence of a cosmic fluid having exotic properties is postulated, the  \textit{dark energy}, or introducing a  cosmological constant which, in turn, brings about other problems, concerning its nature and origin \cite{Peebles:2002gy}. Similarly,  the rotation curves of spiral galaxies seem to escape the current gravity paradigms: to reconcile the theoretical models with the observations, the existence of a peculiar form of matter is required, the \textit{dark matter}, which is supposed to be a cold and pressureless medium, whose distribution is that of a spherical halo around the galaxies\cite{Binney87}. 

Even though in the last 100 years GR has undergone brilliant successes in explaining and discovering the features of the gravitational interactions in the world around us (a comprehensive overview about the tests of GR in the last 100 years can be found in the recent paper by Clifford M. Will \cite{Will:2014bqa}),  the reliability of  Einstein theory  and its principles are questioned when observations on  large cosmological scales are confronted with the theoretical predictions or, also, when the issue of quantization of gravity is considered.  The quest for a more fundamental theory of gravity is open: this new theory must be capable of describing gravitational interactions at all scales; moreover,  since GR is in excellent agreement with the observations at small scales, any new theory of gravity should lead to Einstein theory in some suitable limit. In the recent paper \cite{Berti:2015itd} a review of the motivations to consider extensions of GR can be found, together with  a discussion of some modified theories of gravity. 

A possible  approach toward a new theory of gravity consists in extending GR on a purely geometric basis:  theories of gravity having a richer geometric structure have been proposed, such as the $f(R)$ theories,  where the gravitational Lagrangian depends on a function $f$ of the curvature scalar $R$ (see \cite{Sotiriou:2008rp,Capozziello:2007ec} and references therein): when $f(R)=R$ the action reduces to the usual Einstein-Hilbert action, and Einstein's theory is obtained. Another approach stems from a  generalization of teleparallel gravity (TEGR) \cite{pereira,Aldrovandi:2003xu,Maluf:2013gaa}: the latter is based on a Riemann-Cartan space-time, endowed with the nonsymmetric Weitzenb\"ock connection which, unlike the Levi-Civita connection of GR, gives rise to torsion but it is curvature-free. In TEGR  torsion plays the role of curvature, while the tetrads field plays the role of the dynamical field instead of the metric field; the field equations are obtained from a Lagrangian containing the torsion scalar $T$. It is interesting to point out that Einstein himself studied a theory of gravity with torsion, in  his attempt to formulate a unified theory of gravitation and electromagnetism \cite{einstein1,einstein2,einstein3}. Even though the two theories have a different geometric structure,  GR and TEGR share the same dynamics: this means that  every solution of GR is also solution of TEGR. In the so called $f(T)$ theories, the gravitational Lagrangian is an  analytic function of the torsion scalar $T$,  so they  generalize  TEGR just as the $f(R)$ theories do for GR. These theories are appealing because they are not equivalent to GR  \cite{Ferraro:2008ey, Fiorini:2009ux} and, as a consequence, they can be considered as potential  candidates to solve the issue of cosmic acceleration \cite{cardone12,Myrzakulov:2010vz,Nashed:2014lva,Yang:2010hw,bengo,kazu11,Karami:2013rda,sari11,cai11,capoz11,Bamba:2013jqa,Camera:2013bwa,Nashed:2015pda}. Various aspects of $f(T)$ gravity have been considered, such as, for instance, exact solutions and stellar models  \cite{Wang:2011xf,Ferraro:2011ks,Gonzalez:2011dr,Capozziello:2012zj,Rodrigues:2013ifa,Nashed:uja,Nashed:2015qza,Junior:2015fya,Bejarano:2014bca,ss3,ss4,ss6,ss7}. Spherically symmetric solutions in $f(T)$ gravity are important also because these solutions, describing the gravitational field of pointlike sources, can be used to constrain these theories with planetary motions in the Solar System. A weak-field solution, suitable to model the gravitational interaction in the Solar System,  was obtained by Iorio and Saridakis  (IS) \cite{Iorio12}   for a Lagrangian in the form $f(T)=T+\alpha T^{2}$, where $\alpha$ is a small constant which parametrizes the departure from GR: this solution was used to constrain the $\alpha$ parameter in the Solar System \cite{Xie:2013vua}. Indeed, the additional degrees of freedom of $f(T)$  gravity are related to the fact that the equations of motion are not invariant under local Lorentz transformations \cite{Li:2010cg} (see however the recent papers \cite{Krssak:2015lba,Krssak:2015oua} where the possibility of obtaining a fully covariant reformulation of $f(T)$ gravity has been analyzed).  In particular, this implies the existence of a preferential global reference frame defined by the autoparallel curves of the manifold that solve the equations of motion. Consequently,  even though the symmetry can help in choosing suitable coordinates to write the metric in a simple way, this does not give any hint on the form of the tetrad.  As discussed in  \cite{tamanini12},  a diagonal tetrad is not a good choice to properly parallelize the spacetime  both in the context of non flat homogeneous and isotropic cosmologies (Friedman-Lemaitre-Robertson-Walker universes) and in spherically symmetric space-times (Schwarzschild or Schwarzschild-de Sitter solutions): actually,  the IS solution was obtained using  a diagonal tetrad, hence it suffers from these theoretical conundrums. In a subsequent paper Ruggiero and Radicella (RR) \cite{Ruggiero:2015oka} obtained a new solution for Lagrangian in the general form $f(T)=T+\alpha T^{n}$, with $|n| \neq 1$,  following the approach described in \cite{tamanini12}, which does not constrain the functional form of the Lagrangian.  A preliminary analysis of the impact of the RR solution on the Solar System dynamics was carried out in \cite{Iorio:2015rla}. 

Here, we investigate the application of this spherically symmetric spacetime solution to the four Solar System tests of GR: perihelion precession, light bending, Shapiro time delay and gravitational redshift. By doing so, we investigate the effect of the $\alpha$ parameter to these tests, and by using observational data, constraints on this parameter are set. 

This work is organized as follows. In Sec. II, a brief description of the field equations and solutions for $f(T)$ gravity with a non diagonal tetrad is given, followed by the applications for Solar System tests being, perihelion precession (Sec. III, light bending (Sec. IV), Shapiro time delay (Sec. V) and gravitational redshift (Sec. VI). Finally, a discussion of the results and constraints on the parameter $\alpha$ is given in Sec. VII.

%------------------------Section-------------------------
\section{II. Field equations and solutions}\label{sec:fT-gravity}
%------------------------Section-------------------------

In this section, we first discuss the foundations of $f(T)$ gravity and obtain the field equations, then, we solve the field equations in spherically symmetry and weak-field approximation, in order to obtain the RR solution.  In $f(T)$ gravity the tetrads play the role of the dynamical field instead of the metric:  given a coordinate basis, the components $e^a_\mu$ of the tetrads are related to the metric tensor $g_{\mu\nu}$ by $g_{\mu \nu}(x) = \eta_{a b} e^a_\mu(x) e^b_\nu(x)$, with  $\eta_{a b} = \text{diag}(1,-1,-1,-1)$. We point out that, in our notation,  latin indexes refer to the tangent space, while greek indexes label coordinates on the manifold, and we set units such that $c=1$. The field equations can be obtained by varying the action 
\begin{equation}
{\cal{S}} = \frac{1}{16 \pi G} \int{ f(T)\, e \, d^4x} + {\cal{S}}_M \ ,
\label{eq:action}
\end{equation}

\noindent with respect to the tetrads, where   $e = \text{det} \  e^a_\mu = \sqrt{-\text{det}(g_{\mu \nu})}$ and ${\cal{S}}_M$ is the action for the matter fields. In the action (\ref{eq:action}), $f$ is a differentiable function of the torsion scalar $T$: in particular, if $f(T)=T$, the action is the same as in TEGR, and the theory is equivalent to GR.  Starting from the tetrads, it is possible to define the torsion tensor, 
\beq
T^\lambda_{\ \mu \nu} = e^\lambda_a \left( \partial_\nu e^a_\mu - \partial_\mu e^a_\nu \right ), \  \label{eq:deftorsiont}
\eeq
and the superpotential tensor
\beq
S^\rho_{\ \mu \nu} = \frac{1}{4} \left ( T^{\rho}_{\ \ \mu \nu} - T_{\mu \nu}^{\ \ \rho}+T_{\nu \mu}^{\ \ \rho} \right ) +
\frac{1}{2} \delta^\rho_\mu T_{\sigma \nu}^{\ \ \sigma} - \frac{1}{2} \delta^\rho_\nu T_{\sigma \mu}^{\ \ \sigma},  \label{eq:defsuperpotential}
\eeq
from which it is possible to obtain the torsion scalar
\beq
T = S^\rho_{\ \mu \nu} T_\rho ^{\ \mu \nu}.  \label{eq:deftorsions}
\eeq
By variation of the action (\ref{eq:action}) with respect to the tetrads field $e^a_\mu$, we obtain the field equations
\begin{widetext}
\beq
e^{-1}\partial_\mu(e\  e_a^{\ \rho}   S_{\rho}^{\ \mu\nu})f_T+e_{a}^{\ \lambda} S_{\rho}^{\ \nu\mu} T^{\rho}_{\ \mu\lambda} f_T
+  e_a^{\ \rho}  S_{\rho}^{\ \mu\nu}\partial_\mu (T) f_{TT}+\frac{1}{4}e_a^{\nu} f = 4\pi G e_a^{\ \mu} {\cal{T}}_\mu^\nu,
\label{eq: fieldeqs}
\eeq
\end{widetext}
in terms of the matter-energy tensor ${\cal{T}}^\nu_\mu$;  the subscripts $T$, here and henceforth, denote differentiation with respect to $T$. 

We are interested in spherically symmetric solutions that can be used to describe the gravitational field of a pointlike source, e.g. of the Sun. To this end, we write the space-time metric in the form 
\begin{equation}
ds^2=e^{A(r)}dt^2-e^{B(r)}dr^2-r^2 d\Omega^2 \ , \label{metric-ansatz}
\end{equation}
where $d\Omega^{2}= d\theta^2+ \sin^2 \theta d\phi^2$. According to the approach described in  \cite{tamanini12}, we use the non diagonal tetrad 
$$
e_\mu^a= \scalemath{0.85}{ \left( \begin{array}{cccc}
e^{A/2}         &   0                                             &   0                                            &    0         \\
0                 &e^{B/2} \sin{\theta}\cos{\phi}   & e^{B/2} \sin{\theta}\sin{\phi}&  e^{B/2} \cos{\theta}\\
0                 &-r \cos{\theta}\cos{\phi}   & -r  \cos{\theta}\sin{\phi}&  r \sin{\theta}\\
0                 &r \sin{\theta}\sin{\phi}   & -r  \sin{\theta}\cos{\phi}&  0\\
 \end{array} \right) }
$$
to obtain the field equations. As clearly discussed in \cite{tamanini12}, a diagonal tetrad that gives back the metric in Eq. (\ref{metric-ansatz}) is not a good choice since the equations of motion for such a choice would constrain \textit{a priori} the form of the Lagrangian. This is related to the lack of the local Lorentz invariance of $f(T)$ gravity:  tetrads connected by local Lorentz transformations lead to the same metric - i.e. the same causal structure - but different equations of motions, thus physically inequivalent solutions. 

Consequently, we obtain the following field equations:
\begin{widetext} 
\begin{eqnarray}
&&\frac{f(T)}{4}-f_T\frac{e^{-B (r)}}{4r^2}\left(2-2e^{B(r)}+r^2 e^{B(r)} T -2r B'(r)\right)+\nonumber\\
&&-f_{TT} \frac{T'(r) e^{-B(r)}}{r}\left(1+e^{B(r)/2}\right)=4\pi \rho, \label{00eq1}\\
&&-\frac{f(T)}{4}+f_T\frac{e^{-B (r)}}{4r^2}\left(2-2e^{B(r)}+r^2 e^{B(r)} T -2r A'(r)\right)=4 \pi p, \label{11eq1}\\
&&f_T\left[-4+4e^{B(r)}-2r A'(r)-2r B'(r)+r^2A'(r)^2-r^2A'(r)B'(r)+2r^{2}A''(r)\right]+\nonumber\\
&&+2rf_{TT}T'\left(2+2e^{B(r)/2}+rA'(r)\right)=0, \label{eq31}
\end{eqnarray}
\end{widetext}
where $\rho,\ p$ are the energy density and pressure of the matter energy-momentum tensor; the prime denotes differentiation with respect to the radial coordinate $r$. The expression of the torsion scalar turns out to be
\begin{equation}\label{torsion-scalar-general}
T=\frac{2e^{-B(r)}(1+e^{B(r)/2})}{r^{2}}\left[1+e^{B(r)/2}+r A'(r)\right].
\end{equation}

A thorough discussion of the exact solutions in vacuum ($\rho=p=0$) and in presence of a cosmological constant ($\rho=-p$) of the above field equations can be found in \cite{tamanini12}. Here we consider the  weak-field solutions with nonconstant torsion scalar, i.e. $T'=dT/dr \neq 0$, and we are interested in constraining these solutions by means of Solar System tests: so, we can safely suppose that these solutions are perturbations of a flat background Minkowski space-time. As a consequence, we write $e^{A(r)}=1+A(r), \quad e^{B(r)}=1+B(r)$ for the metric coefficients. Furthermore, in solving the field equations (\ref{00eq1})-(\ref{eq31}) we confine ourselves to linear perturbations, and consider $f(T)$ in the form  $f(T)=T+\alpha T^n$, where $\alpha$ is a small constant, parametrizing the departure of these theories from $GR$, and $|n| \neq 1$. 

On looking for solutions of the equations (\ref{00eq1})-(\ref{eq31}) with $\rho=k$, $p=-k$, which corresponds to a cosmological constant, we obtain the following general expressions
\beq
A(r)=-\frac{C_{1}}{r}-\alpha \frac{r^{2-2n}}{2n-3}2^{3n-1} -\frac 1 3 \Lambda r^{2},
\label{eq:solakn}
\eeq
\beq
B(r)=\frac{C_{1}}{r}+\alpha \frac{r^{2-2n}}{2n-3} 2^{3n-1} \left(-3n+1+2n^{2} \right)+\frac 1 3 \Lambda r^{2},
\label{eq:solbkn}
\eeq
where we set $k=\frac{\Lambda}{8\pi}$ and $\Lambda$ is the cosmological constant. Notice that, on setting $C_{1}=2M$, we obtain a weak-field Schwarzschild - de Sitter solution  perturbed by terms that are proportional to $\alpha$ and decay with a power of the radial coordinate. Moreover, the torsion scalar is
\beq
T(r)=\frac{8}{r^{2}}+2\alpha r^{-2n}2^{3n} \left(n+1 \right). \label{torsion-scalar-metric}
\eeq
We see that the perturbation terms proportional to $\alpha$ go to zero both when $r \rightarrow \infty$ with $n > 1$ and  when $r \rightarrow 0$ with $n<1$\footnote{In the latter case,  in order the keep the perturbative approach self-consistent, a maximum value of $r$ must be defined  to consider these terms as perturbations of the flat space-time background.}.   It is interesting to point out that our linearized approach can be applied to arbitrary polynomial corrections to the torsion scalar. In other words, a  general Lagrangian can be written in the form $f(T)=T+g(T)$, where $g(T)$ can be seen as a perturbation of the TEGR-GR Lagrangian $f(T)=T$. As a consequence, by writing the arbitrary function $g(T)$ as a suitable power series, it is possible to evaluate its impact  as a perturbation of the weak-field spherically symmetric solution in GR: the $n$th term of the series gives a contribution proportional to $r^{2-2n}$. 

The case with $n=2$, corresponding to the Lagrangian $f(T)=T+\alpha T^{2}$ is interesting since every general Lagrangian reduces to this form, in first approximation. Our vacuum ($\Lambda=0$) solution  for $n=2$, turns out to be
\beq
A(r)=-32\,{\frac {\alpha}{{r}^{2}}}-{\frac {{\it C_{1}}}{r}}, \label{eq:sola}
\eeq
\beq
B(r) =96\,{\frac {\alpha}{{r}^{2}}}+{\frac {{\it C_{1}}}{
r}}. \label{eq:solb}
\eeq
Our results can be compared to  the IS solution \cite{Iorio12}, where a Lagrangian in the form $f(T)=T+\alpha T^{2}$ was considered.  While to lowest order approximation in both cases the perturbations are proportional to $1/r^{2}$, the numerical coefficients are different: this is not surprising, since the IS solution was obtained by solving different field equations using a diagonal tetrad.

\section{III. Perihelion Precession}\label{sec:per-precession}

Consider the metric Eq. \eqref{metric-ansatz}, and consider the equatorial plane $\theta = \pi/2$. For timelike orbits, $ds^2 = d\tau^2$, where $\tau$ is the proper time, and hence the metric reduces to
\begin{equation}
1 = e^{A(r)} \: \left(\dfrac{dt}{d\tau}\right)^2 - e^{B(r)} \: \left(\dfrac{dr}{d\tau}\right)^2 - r^2 \: \left(\dfrac{d\phi}{d\tau}\right)^2.
\label{timelike-metric}
\end{equation}

Since teleparallel gravity expresses the gravitational field through torsional stresses, geodesics do not exist in a real sense. Instead, we have force equations analogous to the Lorentz equation of electrodynamics, in which for spinless particles is given by 

\begin{equation}
\frac{d^2x^\mu}{d \lambda^2}+{\widehat{\Gamma}^\mu}_{\nu \rho}\frac{dx^\nu}{d\lambda}\frac{dx^\rho}{d\lambda}={K^\mu}_{\nu \rho}\frac{dx^\nu}{d\lambda} \frac{dx^\rho}{d\lambda},
\label{force_eq}
\end{equation}

\noindent where $\udt{K}{\lambda}{\mu\nu}$ is the contortion tensor defined to be,
\begin{align}
\udt{K}{\lambda}{\mu\nu}&\equiv\udt{\widehat{\Gamma}}{\lambda}{\mu\nu}-\udt{\Gamma}{\lambda}{\mu\nu}\nonumber\\
&=\frac{1}{2}\left(\udt{T}{\lambda}{\mu\nu}+\dut{T}{\mu\nu}{\lambda}+\dut{T}{\nu\mu}{\lambda}\right).
\end{align}

\noindent Through the right-hand side (RHS) of Eq.(\ref{force_eq}) the torsion plays the role of gravitational force.  From the definition of the contortion tensor, the geodesic equation of GR can be derived,
\begin{equation}
\frac{d^2x^\mu}{d\lambda^2}+\Gamma^\mu_{\nu \rho}\frac{dx^\nu}{d\lambda}\frac{dx^\rho}{d \lambda}=0.
\end{equation}

However, keeping in mind that both the geodesic equation and the forcelike equation share a common source, the final trajectories will be the same thus giving the equivalence between general relativity and teleparallel gravity \citep{deAndrade:1997gka,DeAndrade:2000sf,aldrovandi2012teleparallel}. For this reason, using the geodesic equation instead, from the time $t$ and $\phi$ components (i.e. $\mu = 0, 3$), the following constants of motion are obtained,
\begin{align}
e^{A(r)} \dfrac{dt}{d\tau} &= E, \label{energy-timelike} \\
r^2 \dfrac{d\phi}{d\tau} &= L, \label{angular-momentum-timelike}
\end{align}

\noindent where $E$ is the energy per unit mass and $L$ is the angular momentum per unit mass of the particle. It is important to note that this equivalence only exists when the weak equivalence principle (WEP) holds. For the case in which the WEP does not hold, the force equations of teleparallel gravity change, depending on the inertial mass and gravitational mass of the particle, effectively changing the equations to depend on the properties of the test particle itself. This however is consistent within the context of teleparallelism, whilst this is not true with regards to GR. Since the weak-field limit is considered ($C_1 = 2M$), the WEP will be assumed to hold \cite{Aldrovandi:2003xu}.

Substituting these conserved quantities in Eq. \eqref{timelike-metric}, we obtain
\begin{equation}
1 = \dfrac{E^2}{e^{A(r)}} - \dfrac{e^{B(r)} L^2}{r^4} \left(\dfrac{dr}{d\phi}\right)^2 - \dfrac{L^2}{r^2}.
\label{perihelion-DE}
\end{equation}

\noindent Solving for the angle between perihelion $r_-$ and aphelion $r_+$, we get
\begin{align}
&\int\limits_{\phi(r_-)}^{\phi(r_+)} d\phi = \phi(r_+) - \phi(r_-) \nonumber \\
&= \int\limits_{r_-}^{r_+} \dfrac{e^{B(r)/2}}{r^2} \left(\dfrac{E^2}{L^2 e^{A(r)}} - \dfrac{1}{L^2} - \dfrac{1}{r^2} \right)^{-1/2} \: dr, \label{integral-perihelion}
\end{align}

\noindent which leads directly to the total perihelion precession per orbit,
\begin{equation}
\Delta \phi = 2|\phi(r_+)-\phi(r_-)| - 2\pi.
\label{total-perihelion-angle}
\end{equation}

In order to evaluate the integral in Eq. \eqref{integral-perihelion}, we shall make use of properties of elliptical orbits. At $r = r_{\pm}$, we have that $dr/d\phi = 0$. Thus, Eq. \eqref{perihelion-DE} reduces to
\begin{equation}
1 = \dfrac{E^2}{e^{A(r_\pm)}} - \dfrac{L^2}{{r_\pm}^2},
\end{equation}

\noindent which is a system of simultaneous equations for $E^2$ and $L^2$. The solutions are
\begin{align}
E^2 &= \dfrac{e^{A(r_-)} e^{A(r_+)} \left({r_+}^2 - {r_-}^2\right)}{e^{A(r_-)}{r_+}^2 - e^{A(r_+)}{r_-}^2}, \label{E-value} \\
L^2 &= \dfrac{{r_+}^2 {r_-}^2 \left(e^{A(r_-)}-e^{A(r_+)}\right)}{e^{A(r_+)}{r_-}^2-e^{A(r_-)}{r_+}^2}. \label{L-value}
\end{align}

\noindent Thus, Eq. \eqref{integral-perihelion} becomes

\begin{widetext}
\begin{equation}
\phi(r_+) - \phi(r_-) = \int\limits_{r_-}^{r_+} \dfrac{e^{B(r)/2}}{r^2} \left[\dfrac{e^{A(r_-)} e^{A(r_+)} \left({r_+}^2 - {r_-}^2\right) + e^{A(r)}\left(e^{A(r_+)}{r_-}^2-e^{A(r_-)}{r_+}^2\right)}{{r_+}^2 {r_-}^2 \left(e^{A(r_+)}-e^{A(r_-)}\right) e^{A(r)}} - \dfrac{1}{r^2} \right]^{-1/2} \: dr.
\label{integral-perihelion-modified-1}
\end{equation}
\end{widetext}

The square bracketed term in Eq. \eqref{integral-perihelion-modified-1} vanishes when $r = r_{\pm}$. Thus, we have that 
\begin{equation}
\dfrac{E^2}{L^2 e^{A(r)}} - \dfrac{1}{L^2} - \dfrac{1}{r^2} = D(r) \left(\dfrac{1}{r} - \dfrac{1}{r_+}\right) \left(\dfrac{1}{r_-} - \dfrac{1}{r}\right),
\label{D(r)-equation}
\end{equation}

\noindent where $D(r)$ is some function in $r$. Since we are considering slightly eccentric orbits, $D(r)$ is approximately constant, i.e. $D(r) \approx D$. Thus, the value of $D$ can be obtained by evaluating the expression at any point along the orbit. For simplicity, we evaluate the expression at $r = \mathcal{L}$, the \textit{semi-latus rectum}, which is defined to be
\begin{equation}
\dfrac{1}{\mathcal{L}} \equiv \dfrac{1}{2} \left(\dfrac{1}{r_+} + \dfrac{1}{r_-} \right) = \dfrac{1}{a (1-e^2)}.
\label{semi-latus-def}
\end{equation}

\noindent where $a$ is the semimajor axis and $e$ is the eccentricity. 

Using the change of variables, $u = 1/r$, and defining $X(u) \equiv e^{-A(u)}$, differentiating Eq. \eqref{D(r)-equation} yields
\begin{equation}
D = 1 - \dfrac{(u_+ - u_-)(u_+ + u_-)}{2 \left(X(u_+) - X(u_-)\right)} X''(u) \bigg|_{u = \mathcal{L}^{-1}}.
\label{D-equation-1}
\end{equation}

\noindent Since we are only considering slightly elliptic orbits, we have that
\begin{equation}
X(u_+) - X(u_-) \approx (u_+ - u_-) X'(\mathcal{L}^{-1}),
\end{equation}

\noindent which simplifies Eq. \eqref{D-equation-1} to
\begin{equation}
D = 1 - \dfrac{u X''(u)}{X'(u)} \bigg|_{u = \mathcal{L}^{-1}}.
\label{D-equation-2}
\end{equation}

Substituting Eq. \eqref{D(r)-equation} with $D(r) \approx D$ into Eq. \eqref{integral-perihelion}, we get
\begin{align}
&\phi(r_+) - \phi(r_-) \simeq \nonumber \\
& \int\limits_{r_-}^{r_+} \dfrac{e^{B(r)/2}}{r^2} \left[D \left(\dfrac{1}{r} - \dfrac{1}{r_+}\right) \left(\dfrac{1}{r_-} - \dfrac{1}{r}\right) \right]^{-1/2} \: dr.
\end{align}

\noindent Using the following change of variables, $u = 1/r$, and
\begin{equation}
u = \dfrac{1}{2} (u_+ + u_-) + \dfrac{1}{2} (u_+ - u_-) \sin \psi,
\end{equation}

\noindent in succession, the integral becomes,
\begin{equation}
\phi(r_+) - \phi(r_-) = D^{-1/2} \int\limits_{-\pi/2}^{\pi/2} e^{B(\psi)/2} \: d\psi.
\label{integral-perihelion-modified-2}
\end{equation}

The next step is to find $D$ and solve the integral in Eq. \eqref{integral-perihelion-modified-2} for general $n$. In what follows, the result holds for every $n$ except for $n = -1,1,3/2$. Using Eq. \eqref{D-equation-2}, the value of $D$ up to first order in $\alpha$, $\Lambda$ and $M$, and neglecting their products, is found to be
\begin{align}
&D \simeq 1-\frac{4 M}{\mathcal{L}}+\frac{4 \Lambda \mathcal{L}^2}{3}-\frac{\Lambda \mathcal{L}^3}{M} \nonumber \\
&-\frac{\alpha 2^{3 n-1} (n-1) \mathcal{L}^{3-2 n}}{M}-\frac{\alpha 2^{3 n+1} (n-1) \mathcal{L}^{2-2 n}}{2 n-3}.
\end{align}

The integral in Eq. \eqref{integral-perihelion-modified-2} up to first order in $\alpha$, $\Lambda$ and $M$, and neglecting their products, is
\begin{align}
\int\limits_{-\pi/2}^{\pi/2} &e^{B(\psi)/2} \: d\psi \simeq \pi+\frac{\pi  M}{\mathcal{L}}+\frac{\pi  \Lambda \mathcal{L}^2}{6} \nonumber \\
&+ \alpha \frac{\pi 2^{3 n-2} \left(2 n^2-3 n+1\right) \mathcal{L}^{2-2 n}}{2 n-3}.
\end{align}

\noindent Thus, using Eqs. \eqref{total-perihelion-angle} and \eqref{integral-perihelion-modified-2}, as well as the definition of the semi-latus rectum in Eq. \eqref{semi-latus-def}, the total perihelion precession per orbit up to first order in $\alpha$, $\Lambda$ and $M$ is
\begin{align}
\hspace{-2mm} \Delta\phi &\simeq \frac{6 \pi  M}{a(1-e^2)}+\frac{\pi  \Lambda a^3(1-e^2)^3}{M} \nonumber \\
&+\frac{\pi  \alpha 2^{3 n-1} (n-1) a^{3-2 n} (1-e^2)^{3-2 n}}{M} \nonumber \\
&+\frac{\pi \alpha 2^{3 n+1} (n-1) n a^{2-2 n} (1-e^2)^{2-2 n}}{2 n-3}.
\label{perihelion-general-solution}
\end{align}

If $\alpha = 0$, we obtain the Schwarzchild de-Sitter solution as expected. For the case $n = 0$, the following solution is obtained,
\begin{align}
\hspace{-2mm} \Delta\phi &\simeq \frac{6 \pi  M}{a(1-e^2)}+\frac{\pi  \Lambda a^3(1-e^2)^3}{M}-\frac{\pi \alpha a^3 (1-e^2)^3}{2 M}.
\end{align}

\noindent One can note that the $\alpha$ takes the role of a cosmological constant and is of the same form as the de-Sitter term, which makes sense since $n = 0$ reduces to the Schwarzchild de-Sitter like solution with a modified cosmological constant. Using the transformation,
\begin{equation}
\dfrac{\alpha r^2}{6} - \dfrac{\Lambda r^2}{3} \equiv -\dfrac{k r^2}{3},
\label{new-cosmological-constant-n0}
\end{equation}

\noindent with the total perihelion precession angle per orbit for a Schwarzchild de-Sitter solution with $k$ as the cosmological constant Ref. \cite{Sultana:2012qp,Miraghaei:2008vb,rindler2006relativity}, the above solution is obtained . 

On the other hand, for $n=2$ and $n=3$, the following solutions are obtained,
\begin{align}
\hspace{-2mm} \Delta\phi &\simeq \frac{6 \pi  M}{a(1-e^2)}+\frac{\pi  \Lambda a^3(1-e^2)^3}{M}+\frac{256 \pi  \alpha}{a^2(1-e^2)^2} \nonumber \\
&+\frac{32 \pi  \alpha}{a(1-e^2) M}, \label{perihelion-solution-n2} \\
\hspace{-2mm} \Delta\phi &\simeq \frac{6 \pi  M}{a(1-e^2)}+\frac{\pi  \Lambda a^3(1-e^2)^3}{M}+\frac{2048 \pi  \alpha}{a^4(1-e^2)^4}\nonumber \\
&+\frac{512 \pi  \alpha}{a^3(1-e^2)^3 M}. \label{perihelion-solution-n3}
\end{align}

For larger values of integer $n$, the last term in $\alpha$ becomes much smaller than the other term in $\alpha$, and hence can be neglected.

\begin{table*}[t!]
  \centering
    \begin{tabular}{c c c c c}
    \hline
    \rule{0pt}{5ex} $Planet$ & $L$ ($10^6$ km) &  Rev./cty. & Obs. prec. Corr. ($^{\prime\prime}$/cty.) & $\alpha$ ({km}$^2$) \\ [2ex] \hline
    \rule{0pt}{3ex} Mercury   & 55.4430 & 414.9378 & $-$0.0040$\pm$0.0050 & $-8.554 \times 10^{-5}$ $< \alpha <  9.505 \times 10^{-6}$ \\
    Venus   & 108.1947 & 162.6016 & 0.0240$\pm$0.0330 & $-4.260 \times 10^{-4} < \alpha < 2.698 \times 10^{-3}$\\
    Earth  & 149.5568  & 100.0000 & 0.0060$\pm$0.0070 &  $-1.064 \times 10^{-4} < \alpha < 1.383 \times 10^{-3}$\\
    Mars  & 225.9289 & 53.1915 & $-$0.0070$\pm$0.0070 & $-4.230 \times 10^{-3} < \alpha < -8.072 \times 10^{-15}$ \\
    \hline
    \end{tabular}
    \caption{Different values of $\alpha$ for the first four planets for the case $n = 2$ using $\Lambda \sim 10^{-46} \: \text{km}^{-2}$.}
    \label{FT_Solutions_Constraint_n2}
\end{table*}

\begin{table*}[t!]
  \centering
    \begin{tabular}{c c c c c}
    \hline
    \rule{0pt}{5ex} $Planet$ & $L$ ($10^6$ km) &  Rev./cty. & Obs. prec. Corr. ($^{\prime\prime}$/cty.) & $\alpha$ ({km}$^4$) \\ [2ex] \hline
    \rule{0pt}{3ex} Mercury   & 55.4430 & 414.9378 & $-$0.0040$\pm$0.0050 & $-1.643 \times 10^{10} < \alpha < 1.826 \times 10^9$ \\
    Venus   & 108.1947 & 162.6016 & 0.0240$\pm$0.0330 & $-3.117 \times 10^{11} < \alpha < 1.97385 \times 10^{12}$\\
    Earth  & 149.5568  & 100.0000 & 0.0060$\pm$0.0070 &  $-1.487 \times 10^{11} < \alpha < 1.933 \times 10^{12}$\\
    Mars  & 225.9289 & 53.1915 & $-$0.0070$\pm$0.0070 & $-1.349 \times 10^{13} < \alpha < -25.753$ \\
    \hline
    \end{tabular}
    \caption{Different values of $\alpha$ for the first four planets for the case $n = 3$ using $\Lambda \sim 10^{-46} \: \text{km}^{-2}$.}
    \label{FT_Solutions_Constraint_n3}
\end{table*}

One can note that the solution obtained in Eq. \eqref{perihelion-general-solution} gives the Schwarzchild de-Sitter terms. Furthermore, two first order contribution terms in $\alpha$ are found in which their contribution to the perihelion precession is dependent on the sign of $\alpha$. 

Before we determine the resulting contribution, we first rewrite the $\alpha$ terms as,
\begin{equation}
\pi \alpha (n-1) \mathcal{L}^{2-2 n} 2^{3n} \left[\frac{\mathcal{L}}{2M}+\frac{2n}{2 n-3}\right].
\end{equation}

\noindent Recall that we have considered our solution for $n \geq 0$ and excluding the cases $n = 0, 1, 3/2$ (because $n = 0$ is the modified cosmological constant case, $n = 1$ and $n = 3/2$ are not allowed). Since $L >> 1$ for the orbits considered in our Solar System, $\dfrac{3L}{2(L+2M)} \approx \dfrac{3}{2}$. The contributions depending on the sign of $\alpha$ and magnitude of $n$ are as follows,

\begin{enumerate}[label=(\roman*)]
\item $1 < n < \dfrac{3L}{2(L+2M)}$ or $n > \dfrac{3}{2}$: for such cases, a negative $\alpha$ yields a negative contribution whilst a positive $\alpha$ gives a positive contribution.
\item $n = \dfrac{3L}{2(L+2M)}$: in this case, the square bracketed term is exactly $0$. Therefore, there is no $\alpha$ contribution to perihelion precession.
\item $0 < n < 1$ or $\dfrac{3L}{2(L+2M)} < n < \dfrac{3}{2}$: within this range, a negative $\alpha$ yields a positive contribution whilst a positive $\alpha$ gives a negative contribution.
\end{enumerate}

Using the data available in Refs. \citep{brown2015reflrections, IAU:6911132}, the parameter $\alpha$ is constrained using $\Lambda \sim 10^{-46} \: \text{km}^{-2}$ \citep{Carmeli:2000cf} as shown in Tables \ref{FT_Solutions_Constraint_n2} and \ref{FT_Solutions_Constraint_n3}. From the tables, we can set the following constraints. For $n = 2$, we found that 

\begin{equation}
-8.554 \times 10^{-5} < \alpha < -8.072 \times 10^{-15} \: \text{km}^{2},
\label{perihelion-alpha-contrainst-n2}
\end{equation}

\noindent whilst for $n = 3$, we obtained
\begin{equation}
-1.643 \times 10^{10} < \alpha < -25.753 \: \text{km}^{4}.
\label{perihelion-alpha-contrainst-n3}
\end{equation}

\section{IV. Null Orbits and Light Bending}\label{sec:light-bending}

Consider the metric Eq. \eqref{metric-ansatz}, and consider the equatorial plane $\theta = \pi/2$. For null orbits, $ds^2 = 0$, and hence the metric reduces to
\begin{equation}
0 = e^{A(r)} \: \left(\dfrac{dt}{d\lambda}\right)^2 - e^{B(r)} \: \left(\dfrac{dr}{d\lambda}\right)^2 - r^2 \: \left(\dfrac{d\phi}{d\lambda}\right)^2,
\label{null-metric}
\end{equation}

\noindent where $\lambda$ is an affine parameter. Similar to the previous section, we note two constants of motion,
\begin{align}
e^{A(r)} \dfrac{dt}{d\lambda} &= E, \label{energy-null} \\
r^2 \dfrac{d\phi}{d\lambda} &= L, \label{angular-momentum-null}
\end{align}

\noindent where $E$ is the energy and $L$ is the angular momentum. Substituting these conserved quantities in Eq. \eqref{null-metric}, we obtain
\begin{equation}
0 = \dfrac{E^2}{L^2} - \dfrac{e^{A(r)+B(r)}}{r^4} \left(\dfrac{dr}{d\phi}\right)^2 - \dfrac{e^{A(r)}}{r^2}.
\label{light-bending-1st-order-DE}
\end{equation}

Defining a new variable $u = 1/r$ and differentiating the previous expression yields the light bending differential equation,
\begin{align}
0 = \dfrac{d^2 u}{d\phi^2} &+ \dfrac{1}{2} \dfrac{d}{du} \left[A(u)+B(u) \right] \left(\dfrac{du}{d\phi}\right)^2 \nonumber \\
&+ \dfrac{1}{2 e^{A(u)+B(u)}} \dfrac{d}{du} \left[u^2 e^{A(u)}\right]. \label{light-bending-2nd-order-DE}
\end{align}

\noindent To eliminate the square derivative term, we again make use of Eq. \eqref{light-bending-1st-order-DE} and the fact that $E^2/L^2 \equiv 1/R^2$, the impact parameter $R$, to get
\begin{equation}
\left(\dfrac{du}{d\phi}\right)^2 = \dfrac{{u_R}^2- u^2 e^{A(u)}}{e^{A(u)+B(u)}}.
\label{derivative-square-value}
\end{equation}

\noindent where $u_R \equiv 1/R$. Thus, substituting Eq. \eqref{derivative-square-value} in Eq. \eqref{light-bending-2nd-order-DE} gives the final light bending equation,
\begin{align}
0 &= \dfrac{d^2 u}{d\phi^2} + \dfrac{1}{2 e^{A(u)+B(u)}} \dfrac{d}{du} \left[u^2 e^{A(u)}\right] \nonumber \\
&+ \dfrac{1}{2} \dfrac{d}{du} \left[A(u)+B(u) \right] \dfrac{{u_R}^2- u^2 e^{A(u)}}{e^{A(u)+B(u)}}. 
\end{align}

Since the only terms of interest are those up to first order, the differential equation simplifies as follows. Expanding the metric terms, as long as $|A(u)|, |B(u)| < 1$ and since $A(u), B(u)$ are first order functions, we get
\begin{equation}
e^{-A(u)-B(u)} \approx 1 - A(u) - B(u).
\end{equation}

\noindent Thus, since any products of $A(u)$ and $B(u)$ are at least second order, the differential equation reduces to,
\begin{align}
0 \approx &\dfrac{d^2 u}{d\phi^2} + u\left(1 -B(u)\right)+ \dfrac{1}{2} {u_R}^2  \dfrac{d}{du} \left[A(u)+B(u) \right] \nonumber \\
& - \dfrac{1}{2} u^2 \dfrac{d}{du} \left[B(u)\right], 
\end{align}

\noindent which, for the functions of $A(u)$ and $B(u)$ given in the second section, we get
\begin{align}
0 \approx &\dfrac{d^2 u}{d\phi^2} + u - 3Mu^2 +\frac{\alpha 2^{3 n-1} (n-1) n u^{2 n-3}}{R^2} \nonumber \\
&+\frac{\alpha 2^{3 n-1} n \left(3n-2 n^2-1\right) u^{2 n-1}}{2 n-3}. \label{final-light-bending-2nd-order-DE}
\end{align}

\begin{figure}[b]
\begin{center}
\includegraphics[width=0.45\textwidth]{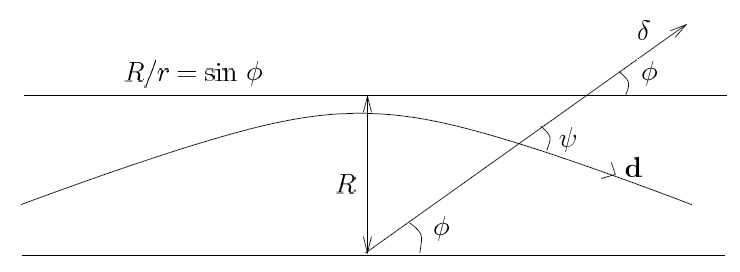}
\end{center}
\caption{The plane graph representing the orbit with the one-sided deflection angle $\epsilon=\psi+\phi$ \citep{Sultana:2013}.}
\label{light-deflection-plot}
\end{figure}

From this differential equation, we obtain a solution for $u$, which represents the path of the photon as a function of $\phi$. Since this is a nonlinear differential equation which is very difficult to solve, we shall adopt Bodenner and Will's iterative method \cite{Bodenner:2003} and then make use of Rindler and Ishak's invariant cosine technique \citep{Rindler:2007zz}, which in turn allows us to determine the deflection angle $\epsilon$ as shown in Fig.\ref{light-deflection-plot}. 

This is achieved by considering a photon coming from a far away source at an angle (without loss of generality) $\phi = -\pi/2$ (but not $r \rightarrow \infty$ due to the presence of the de Sitter horizon), reaches a point of closest approach at $\phi = 0$ and leaving at an angle $\phi = \pi/2$. In this way, a solution for $u$ is achieved. Before moving forward, a transformation $\phi \rightarrow \pi/2 - \phi$ is applied to make the calculations simpler. 

In order to calculate the deflection angle, we consider two coordinate directions $d = (dr,\: d\phi)$ and $\delta = (\delta r,\: 0)$  which represent the direction of the orbit and the direction of the coordinate line $\phi = constant$ respectively. Lastly, we let $\psi$ be the angle between these two coordinate directions, as illustrated in Fig. \ref{light-deflection-plot}. For simplicity, the paths illustrated in the figure are the homogeneous straight line path and the actual path for a Schwarzchild de-Sitter metric (i.e. $\alpha = 0$). 

The angle $\psi$ between the two coordinate directions $d$ and $\delta$ is given by the invariant formula
\begin{equation}
\cos \psi = \dfrac{g_{ij} d^i \delta^j}{\sqrt{g_{ij} d^i d^j} \sqrt{g_{ij} \delta^i \delta^j}}.
\label{invariant-cosine}
\end{equation}

\noindent where $g_{ij}$ is the metric tensor of the subspace. Thus, substituting the metric tensor into this relation yields
\begin{equation}
\cos \psi = \dfrac{|C|}{\left[C^2 + r^2 e^{-B(r)} \right]^{1/2}},
\end{equation}

\noindent where $C \equiv \dfrac{dr}{d\phi}$. Equivalently, the expression can be rewritten as
\begin{equation}
\tan \psi = \dfrac{r e^{-B(r)/2}}{|C|}.
\end{equation}

\noindent Since the angle is very small, the small angle approximation can be used, giving $\psi$ as
\begin{equation}
\psi \approx \dfrac{r e^{-B(r)/2}}{|C|}.
\label{Psi-Approx}
\end{equation}

To obtain the deflection, we evaluate it when $\phi = 0$. Hence, the one-sided deflection will be $\epsilon = \psi_0$, where $\psi_0$ denotes the evaluation of $\psi$ at $\phi = \phi_0$. Denoting the value of $r$ and $C$ at $\phi = \phi_0$ by $r_0$ and $C_0$ respectively, using Eq.(\ref{Psi-Approx}), the one-sided deflection angle is

\begin{widetext}
\begin{equation}
\psi_0 = \dfrac{r_0}{|C_0| \sqrt{1+\dfrac{2M}{r_0} + \alpha \dfrac{{r_0}^{2-2n} \: 2^{3n-1}}{2n-3} (2n^2-3n+1) + \dfrac{\Lambda {r_0}^2}{3}}}.
\label{one-sided-deflection}
\end{equation}
\end{widetext}

\subsection{A. The case $n = 0$}

For the case $n = 0$, keeping up to first order in $\alpha$, $\Lambda$ and $M$, and neglecting their products, Eq. \eqref{final-light-bending-2nd-order-DE} reduces to,
\begin{equation}
0 \simeq \dfrac{d^2 u}{d\phi^2}+u-3 M u^2.
\end{equation}

\noindent Recall that $n = 0$ corresponds to a Schwarzchild de-Sitter like solution with $\alpha$ taking the role of a cosmological constant. One can note that we do not have a contribution from the cosmological constant, as expected when working with a Schwarzchild de-Sitter metric \cite{Rindler:2007zz,Bhadra:2010jr,Ishak:2007ea}. 

\subsection{B. The case $n = 2$}

Consider a quadratic $T$ in $f(T)$ gravity, which corresponds to $n = 2$. Using Eq. \eqref{final-light-bending-2nd-order-DE}, it reduces to,
\begin{equation}\label{DE-n2}
0 = \dfrac{d^2 u}{d\phi^2}+u-3 M u^2+\frac{64 \alpha u}{R^2}-192 \alpha u^3.
\end{equation}

\noindent Assuming a solution of the form $u \approx u_0 + u_1$, we obtain the following system of differential equations,
\begin{align}
&\dfrac{d^2 u_0}{d\phi^2}+u_0 =0, \label{DE-system_n2-1} \\
&\dfrac{d^2 u_1}{d\phi^2}+u_1-3 M {u_0}^2+\frac{64 \alpha}{R^2}u_0-192 \alpha {u_0}^3 =0. \label{DE-system_n2-2}
\end{align}

\noindent Together, these give the solution for $u$, which is 
\begin{align}
u = &\frac{3 M}{2 R^2}+\frac{\cos \phi}{R}-\frac{M \cos (2 \phi)}{2 R^2}+\frac{40 \alpha \phi \sin \phi}{R^3}\nonumber \\
&+\frac{26 \alpha \cos \phi}{R^3}-\frac{6 \alpha \cos (3 \phi)}{R^3}. \label{u-soln_n2}
\end{align}

The solution of $u$ is symmetric about $\phi = 0$, which is expected from the spherical symmetry. Before deriving the properties and deflection, we first change $\phi \rightarrow \pi/2 - \phi$ to obtain,
\begin{align}
u = &\frac{3 M}{2 R^2}+\frac{\sin \phi}{R}+\frac{M \cos (2 \phi)}{2 R^2}+\frac{26 \alpha \sin \phi}{R^3}\nonumber \\
&+\frac{6 \alpha \sin (3 \phi)}{R^3}+\frac{20 \pi \alpha \cos \phi}{R^3}-\frac{40 \alpha \phi \cos \phi}{R^3}.
\end{align}

From this solution of $u$, one can find the value of the distance of closest approach $\bar{r}$ by setting $u(\pi/2) = 1/\bar{r}$. To obtain the deflection, we evaluate the value for $r_0$
\begin{equation}
\dfrac{1}{r_0} = \frac{2 M}{R^2}+\frac{20 \pi \alpha}{R^3},
\end{equation}

\noindent and the value for $C_0$
\begin{align}
C_0 &= -{r_0}^2 \left(\frac{1}{R}+\frac{4 \alpha}{R^3}\right).
\end{align}

\noindent Using these values and substituting into Eq. \eqref{one-sided-deflection}, the one-sided deflection up to first order in $\alpha$, $\Lambda$ and $M$, and neglecting their products, is found to be
\begin{equation}
\psi_0 = \frac{2 M}{R}\left(1+\frac{10 \pi \alpha}{M R}-\frac{\Lambda R^4}{24 M^2}\right).
\label{one-sided-deflection_n2}
\end{equation}

The contribution of $\alpha$ obtained here is in agreement with Ref. \cite{Ruggiero:2016iaq}. One can also note that the $\alpha$ contribution is independent of the source $M$.

\subsection{C. The case $n = 3$}

For a cubic $T$ in $f(T)$ gravity, i.e. $n = 3$, Eq. \eqref{final-light-bending-2nd-order-DE} becomes,
\begin{equation}\label{DE-n3}
0 \simeq \dfrac{d^2 u}{d\phi^2}+u-3 M u^2+\frac{1536 \alpha}{R^2}u^3-2560 \alpha u^5.
\end{equation}

\noindent Assuming a solution of the form $u \approx u_0 + u_1$, we obtain the following system of differential equations,
\begin{align}
&\dfrac{d^2 u_0}{d\phi^2}+u_0=0, \label{DE-system_n3-1} \\
&\dfrac{d^2 u_1}{d\phi^2}+u_1-3 M {u_0}^2+\frac{1536 \alpha}{R^2}{u_0}^3 \nonumber \\
&-2560 \alpha {u_0}^5 = 0. \label{DE-system_n3-2}
\end{align}

\noindent Solving the system of equations, the solution for $u$ is given to be 
\begin{align}
u = &\frac{3 M}{2 R^2}+\frac{\cos \phi}{R}-\frac{M \cos (2 \phi)}{2 R^2}+\frac{224 \alpha \phi \sin \phi}{R^5} \nonumber \\
&+\frac{148 \alpha \cos \phi}{R^5}-\frac{52 \alpha \cos (3 \phi)}{R^5}-\frac{20 \alpha \cos (5 \phi)}{3 R^5}. \label{u-soln_n3}
\end{align}

Again, the solution of $u$ is symmetric about $\phi = 0$, which is expected from the spherical symmetry. Similarly to what was done to the $n = 2$ case, we first change $\phi \rightarrow \pi/2 - \phi$ to obtain,
\begin{align}
u = &\frac{3 M}{2 R^2}+\frac{\sin \phi}{R}+\frac{M \cos (2 \phi)}{2 R^2}+\frac{148 \alpha \sin \phi}{R^5}\nonumber \\
&+\frac{52 \alpha \sin (3 \phi)}{R^5}-\frac{20 \alpha \sin (5 \phi)}{3 R^5}+\frac{112 \pi \alpha \cos \phi}{R^5}\nonumber \\
&-\frac{224 \alpha \phi \cos (\phi)}{R^5}.
\end{align}

The distance of closest approach $\bar{r}$ can be found by setting $u(\pi/2) = 1/\bar{r}$. Obtained a solution for $u$, the value for $r_0$ is
\begin{equation}
\dfrac{1}{r_0} = \frac{2 M}{R^2}+\frac{20 \pi \alpha}{R^3},
\end{equation}

\noindent whilst the value for $C_0$ is
\begin{align}
C_0 &= -{r_0}^2 \left(\frac{1}{R}+\frac{140 \alpha}{3R^5}\right).
\end{align}

\noindent Using these values and substituting into Eq. \eqref{one-sided-deflection}, the one-sided deflection up to first order in $\alpha$, $\Lambda$ and $M$, and neglecting their products, is found to be
\begin{equation}
\psi_0 = \frac{2 M}{R}\left(1+\frac{56 \pi \alpha}{M R^3}-\frac{\Lambda R^4}{24 M^2}\right). \label{one-sided-deflection_n3}
\end{equation}

Similarly to the previous case, the $\alpha$ contribution is independent of the source $M$.

\subsection{D. The case $n \geq 4$}

For such cases, since the powers of $u$ in Eq. \eqref{final-light-bending-2nd-order-DE} are positive, assuming a perturbed solution of the form $u \approx u_0 + u_1$, the following system of equations is obtained,
\begin{align}
&\dfrac{d^2 u_0}{d\phi^2}+u_0=0, \label{DE-system_n-1} \\
&\dfrac{d^2 u_1}{d\phi^2}+u_1-3 M {u_0}^2+\frac{\alpha 2^{3 n-1} (n-1) n {u_0}^{2 n-3}}{R^2} \nonumber \\
&+\frac{\alpha 2^{3 n-1} n \left(3n-2 n^2-1\right) {u_0}^{2 n-1}}{2 n-3} = 0. \label{DE-system_n-2}
\end{align}

Solving this system of equations generates a solution in terms of the hypergeometric function $_2F_1(a,b;c;z)$, and the method becomes unusable. Instead, the following method is applied. The solution of Eq. \eqref{DE-system_n-2} is an addition of particular solutions (by the principle of superposition), and thus the solution for $u_1$ can be expressed as an addition of two particular solutions: the first for $3M {u_0}^2$ (which gives the Schwarzchild contribution), and the second for the $\alpha$ term. 

The first particular solution is,
\begin{equation}
\frac{3 M}{2 R^2}-\frac{M \cos (2 \phi)}{2 R^2},
\end{equation} 
whilst the second for $\alpha$ is in terms of hypergeometric functions. To avoid working with the latter, define the function $g(\phi) \equiv \alpha 2^{3n-1} n f(\phi)$ to be the solution for this particular solution. In other words, the solution for $u_1$ is given by,
\begin{equation}
u_1 = \frac{3 M}{2 R^2}-\frac{M \cos (2 \phi)}{2 R^2} + \alpha 2^{3n-1} n f(\phi).
\end{equation}

\noindent Substituting into Eq. \eqref{DE-system_n-2}, we get
\begin{align}
\dfrac{d^2f}{d\phi^2}+f-&\frac{(n-1)}{2 n-3} R^{1-2n} \cos^{2n-3} \phi \big[-2 n+3 \nonumber \\
&+(2 n-1) \cos ^2 \phi\big]=0,
\label{f-ODE}
\end{align}

\noindent where the solution for $u_0$ is used. From this differential equation, only the particular solution is required. One can note that the solution of $f$ is independent of the source, $M$. Furthermore, through rescaling of the function $f$, one can note that $f(\phi) = \frac{(n-1)}{2 n-3} R^{1-2n} h(\phi)$, where $h$ satisfies,
\begin{align}
\dfrac{d^2h}{d\phi^2}+h-&\cos^{2n-3} \phi \big[-2 n+3 \nonumber \\
&+(2 n-1) \cos ^2 \phi\big]=0.
\label{h-ODE}
\end{align}

Note that if $h(\phi)$ is a solution, then so is $h(-\phi)$ (since they both satisfy the same differential equation). Since the coefficients of $h$ are continuous, and the cosine function is also continuous (and hence the final term is continuous) on the interval considered $\left(\phi \in \left[-\dfrac{\pi}{2},\dfrac{\pi}{2}\right]\right)$, and the homogeneous solution is unique (it is equal to zero since we are only interested in the particular solution), the solution is unique \cite{university2009differential}. Since both $h(\phi)$ and $h(-\phi)$ are solutions, then $h(\phi) = h(-\phi)$, hence an even function (and therefore, so are $f$ and $g$). 

%Thus, since the differential equation is linear, the even function $\dfrac{h(\phi) + h(-\phi)}{2}$ and the odd function $\dfrac{h(\phi) - h(-\phi)}{2}$ are also solutions to the differential equation. Since the particular solution is unique, and the solutions coincide for $\phi = 0$ ($h(0) = h(-0)$), then this leads to $h(\phi) = h(-\phi)$ i.e. $h$ is an even function (and hence $f$) about $\phi = 0$. For general $n$, the solution for this differential equation is in terms of hypergeometric functions. However, for specific values of $n$, the differential equation is solvable and generates a series of cosine and sine terms. This will prove useful later on.

Using this formalism, the solution for $u$ is given to be 
\begin{align}
u = &\frac{3 M}{2 R^2}+\frac{\cos \phi}{R}-\frac{M \cos (2 \phi)}{2 R^2} + \alpha 2^{3 n-1} n f(\phi). \label{u-soln_n}
\end{align}

Again, the solution of $u$ is symmetric about $\phi = 0$ (since $f$ is even), which is expected from the spherical symmetry. Similarly to what was done to the previous cases, we first change $\phi \rightarrow \pi/2 - \phi$ to obtain,
\begin{align}
u = &\frac{3 M}{2 R^2}+\frac{\sin \phi}{R}+\frac{M \cos (2 \phi)}{2 R^2} + \alpha 2^{3 n-1} n f\left(\dfrac{\pi}{2}-\phi\right).
\end{align}

The distance of closest approach $\bar{r}$ can be found by setting $u(\pi/2) = 1/\bar{r}$. Obtained a solution for $u$, the value for $r_0$ is
\begin{equation}
\dfrac{1}{r_0} = \frac{2 M}{R^2}+\alpha 2^{3 n-1} n f\left(\dfrac{\pi}{2}\right),
\end{equation}

\noindent whilst the value for $C_0$ is
\begin{align}
C_0 &= -{r_0}^2 \left(\frac{1}{R}+\alpha 2^{3 n-1} n f^{\prime}\left(\dfrac{\pi}{2}\right)\right).
\end{align}

\noindent Using these values and substituting into Eq. \eqref{one-sided-deflection}, the one-sided deflection up to first order in $\alpha$, $\Lambda$ and $M$, and neglecting their products, is found to be
\begin{align}
\psi_0 = &\frac{2 M}{R}\Bigg(1-\frac{\Lambda R^4}{24 M^2}+\alpha 2^{3n-2} \dfrac{R^2}{M} \Bigg[n f\left(\frac{\pi }{2}\right) \nonumber \\
&-\frac{2^{2n-2} (n-1) (2 n-1) M^{2n-1}}{(2 n-3) R^{4n-2}}\Bigg]\Bigg). \label{one-sided-deflection_n}
\end{align}

\noindent For $n \geq 4$, the second term in the square brackets leads to a higher order term, and therefore is neglected. Hence, the solution reduces to
\begin{align}
\psi_0 = &\frac{2 M}{R}\left(1-\frac{\Lambda R^4}{24 M^2}+\alpha 2^{3n-2} n \dfrac{R^2}{M} f\left(\frac{\pi}{2}\right)\right), 
\end{align}
or in terms of $h$, 
\begin{align}
\psi_0 = &\frac{2 M}{R}\Bigg(1-\frac{\Lambda R^4}{24 M^2} \nonumber \\
&+\alpha 2^{3n-2} n \dfrac{R^{3-2n}}{M} \frac{(n-1) n}{2 n-3} h\left(\dfrac{\pi}{2}\right) \Bigg). \label{one-sided-deflection_ngeq4}
\end{align}
One can again note that even for $n \geq 4$, the $\alpha$ term is independent of the source $M$. 

The method used here is also valid for other values of $n$. From Eq. \eqref{one-sided-deflection_n}, setting $n = 0$ gives the de-Sitter solution [$g = 0$ in this case thus Eq. \eqref{f-ODE} does not exist], and the solutions for $n = 2$ and 3 can be obtained from Eq. \eqref{h-ODE} and Eq. \eqref{one-sided-deflection_ngeq4} [since the assumption used to derive Eq. \eqref{one-sided-deflection_ngeq4} still holds]. Furthermore, since the solution for $u$ is symmetric about $\phi = 0$ (or $\phi = \pi/2$ after transforming), the total deflection is twice the one-sided deflection (this also holds for the other cases of $n$). 

The major downside of this solution is that the coefficient of this contribution cannot be obtained explicitly unless the solution for $h$ is found using Eq. \eqref{h-ODE}, which can be done as long as specific values of $n$ are considered. 

\subsection{E. $\alpha$ constraints}

For $n = 2$ and $n = 3$, a positive $\alpha$ increases the deflection whilst a negative one decreases the deflection. However, nothing can be said for higher values of $n$ since the signature of $h$ is not determined. Using these results, a constraint on $\alpha$ can be set by taking observational values of the light deflection from the Sun. These in turn give a measure of the $\gamma$ PPN parameter which measures the deviation from GR. In other words, we set
\begin{equation}
2 \psi_0 =\gamma \dfrac{4M}{R}.
\label{Deflection_PPN}
\end{equation}

As an approximation, we will take $R$ to be the solar radius. From the analysis of the data obtained by very long baseline interferometry in Ref.\citep{Lambert:2011}, a value of $\gamma = 0.99992 \pm 0.00012$ was obtained. Taking $\Lambda \sim 10^{-46} \: \text{km}^{-2}$, we obtain 
\begin{equation}
-6.533 < \alpha < 1.307 \: \text{km}^{2},
\label{light_bending-alpha-contrainst-n2}
\end{equation}
for $n = 2$, whilst for $n = 3$, we obtain
\begin{equation}
-5.646 \times 10^{11} < \alpha < 1.129 \times 10^{11} \hspace{1mm} \text{km}^{4}.
\label{light_bending-alpha-contrainst-n3}
\end{equation}

\section{V. Shapiro Time Delay}\label{sec:time-delay}

In order to obtain an expression relating the path of the photon and the coordinate time, we use Eqs. \eqref{null-metric}, \eqref{energy-null} and \eqref{angular-momentum-null} to obtain
\begin{equation}
0 = 1 - \dfrac{e^{B(r)}}{e^{A(r)}} \left(\dfrac{dr}{dt}\right)^2 - \dfrac{e^{A(r)} b^2}{r^2},
\label{time-delay-DE}
\end{equation}

\noindent where $b = L/E$ is the \textit{impact parameter}. At the point of closest approach $r = R$, we have $dr/dt = 0$. Thus, the previous equation reduces to
\begin{equation}
b^2 = \dfrac{R^2}{e^{A(R)}}.
\label{impact-parameter-value-2}
\end{equation} 

\noindent Substituting this value for the impact parameter in Eq. \eqref{time-delay-DE} and integrating to find the time between $r$ and $R$ yields,
\begin{align}
t(r,R) &= \int\limits_{t(R)}^{t(r)} dt \nonumber \\
&= \int\limits_{R}^{r} \left[\dfrac{e^{B(r^\prime)-A(r^\prime)}}{1-\frac{R^2 e^{A(r^\prime)-A(R)}}{{r^\prime}^2}} \right]^{1/2} dr^\prime.
\end{align}

\noindent where the substitution $r^\prime = r$ is used in order to differentiate from the coordinate $r$ in the integral expression. Up to first order in $\alpha$, $\Lambda$ and $M$, and neglecting their products, the integral reduces to

\begin{widetext}
\begin{align}
t(r,R) &= \int\limits_{R}^{r} \sqrt{1-\frac{R^2}{{r^\prime}^2}} \Bigg[1+\frac{2 M}{r^\prime}+\frac{M R}{{r^\prime}^2+r R}+\frac{\Lambda {r^\prime}^2}{3}-\frac{\Lambda R^2}{6} \nonumber \\
&+\alpha \frac{2^{3 n-2} \left[R^4 {r^\prime}^{2 n}+ {r^\prime}^2 R^{2 n} \left((2 n^2-3n+2) {r^\prime}^2-(2 n^2-3n+3) R^2\right)\right]}{{r^\prime}^{2 n} R^{2 n} (2 n-3) ({r^\prime}^2-R^2)} \Bigg] \: dr^\prime,
\end{align}

\noindent which reduces to,
\begin{align}
t(r,R) &= \sqrt{r^2-R^2} +\frac{M \sqrt{r^2-R^2}}{r+R}+2 M \ln \left(\frac{r+\sqrt{r^2-R^2}}{R}\right) + \frac{\Lambda}{18} \sqrt{r^2-R^2} \left(R^2+2 r^2\right) \nonumber \\
&+\int\limits_{R}^{r} \sqrt{1-\frac{R^2}{{r^\prime}^2}} \left[\alpha \frac{2^{3 n-2} \left[R^4 {r^\prime}^{2 n}+ {r^\prime}^2 R^{2 n} \left((2 n^2-3n+2) {r^\prime}^2-(2 n^2-3n+3) R^2\right)\right]}{{r^\prime}^{2 n} R^{2 n} (2 n-3) ({r^\prime}^2-R^2)} \right] \: dr^\prime, \label{t_r_R-expression}
\end{align}
\end{widetext}

\noindent For simplicity, we shall denote the integral term  in the previous expression as $\mathcal{T}_n(r,R)$. 

Now consider a signal traveling from Earth to Mercury and back. The delay is maximum when Mercury is at superior conjunction and the signal passes very close to the sun. Thus, by setting the distance of closest approach $R \simeq {R_{\odot}}$, the total maximum delay is found to be, 
\begin{align}
(\Delta t)_{max} = \: &2 \Big[t \left(r_{\oplus}, {R_{\odot}} \right) + t \left(r_{\mercury}, {R_{\odot}} \right) \nonumber \\
&- \sqrt{{r_{\oplus}}^2-{{R_{\odot}}^2}} - \sqrt{{r_{\mercury}}^2-{{R_{\odot}}^2}} \Big]. \label{shapiro-time-delay_general}
\end{align}

\noindent Using Eq. \eqref{t_r_R-expression} in Eq. \eqref{shapiro-time-delay_general}, we get
\begin{widetext}
\begin{align}
(\Delta t)_{max} &= 4 M \ln \left[\frac{\left(\sqrt{{r_{\oplus}}^2-{R_{\odot}}^2}+{r_{\oplus}} \right)\left(\sqrt{{r_{\mercury}}^2-{R_{\odot}}^2}+{r_{\mercury}} \right)}{{R_{\odot}}^2} \right] \nonumber \\
&+\frac{\Lambda}{9} \left[\sqrt{{r_{\oplus}}^2-{R_{\odot}}^2} \left({R_{\odot}}^2+2 {r_{\oplus}}^2\right)+\sqrt{{r_{\mercury}}^2-{R_{\odot}}^2} \left({R_{\odot}}^2+2 {r_{\mercury}}^2\right) \right] \nonumber \\
&+2M \left[\frac{\sqrt{{r_{\oplus}}^2-{R_{\odot}}^2}}{{R_{\odot}}+{r_{\oplus}}}+\frac{\sqrt{{r_{\mercury}}^2-{R_{\odot}}^2}}{{R_{\odot}}+{r_{\mercury}}} \right]+2 \left[ \mathcal{T}_n \left(r_{\oplus}, {{R_{\odot}}} \right) +  \mathcal{T}_n \left(r_{\mercury}, {R_{\odot}} \right) \right].
\end{align}
\end{widetext}

\noindent Now, using the fact that $R_{\odot}/r_{\oplus} << 1$ and $R_{\odot}/r_{\mercury} << 1$, the above expression simplifies to,
\begin{align}
(\Delta t)_{max} &\simeq 4M \left[1 + \ln \left(\frac{4 r_{\oplus} r_{\mercury}}{{R_{\odot}}^2} \right) \right] +\frac{2 \Lambda}{9} \bigg( {r_{\oplus}}^3 \nonumber \\
&+ {r_{\mercury}}^3  \bigg)+2 \left[ \mathcal{T}_n \left(r_{\oplus}, {{R_{\odot}}} \right) +  \mathcal{T}_n \left(r_{\mercury}, {R_{\odot}} \right) \right].
\label{shapiro-time-delay-general_2}
\end{align}

One can note the first two terms in the expression, the first leading in $M$ is the Schwarzchild contribution whilst the second leading in $\Lambda$ is the de-Sitter contribution, as expected \cite{Kagramanova:2006ax,Ishak:2008ex}. The $\mathcal{T}$ terms are all dependent on $\alpha$, and hence will not contribute to the Schwarzchild de-Sitter contributions in Shapiro time delay (except when $n = 0$). Since the integral for $\mathcal{T}(r,R)$ cannot be evaluated for general $n$, we shall consider some specific cases. 

For the case when $n = 0$, we get 
\begin{equation}
\mathcal{T}_0(r,R) = -\frac{\alpha}{36} \sqrt{{r}^2-R^2} \left(R^2+2 {r}^2\right),
\end{equation}

\noindent which is of the same form as the $\Lambda$ term found in Eq. \eqref{t_r_R-expression}. This makes sense since for $n = 0$, $\alpha$ takes role of a cosmological constant. 

Now, for the case when $n = 2$, the integral evaluates to
\begin{equation}
\mathcal{T}_2(r,R) = \frac{80 \alpha \cos ^{-1}\left(R/r\right)}{R}.
\end{equation}

\noindent Using Eq. \eqref{shapiro-time-delay-general_2}, again using the fact that $R_{\odot}/r_{\oplus} << 1$ and $R_{\odot}/r_{\mercury} << 1$, the time delay becomes
\begin{align}
(\Delta t)_{max} &\simeq 4M \left[1 + \ln \left(\frac{4 r_{\oplus} r_{\mercury}}{{R_{\odot}}^2} \right) \right] +\frac{2 \Lambda}{9} \bigg( {r_{\oplus}}^3 \nonumber \\
&+ {r_{\mercury}}^3  \bigg) + 160 \alpha \left(\frac{\pi }{R_{\odot}}-\frac{r_{\oplus}+r_{\mercury}}{r_{\oplus} r_{\mercury}}\right)].
\label{shapiro-time-delay_n=2_solution}
\end{align}

Lastly, for the case when $n = 3$, we obtain 
\begin{equation}
\mathcal{T}_3(r,R) = \frac{896 \alpha \cos ^{-1}\left(R/r\right)}{3 R^3}+\frac{256 \alpha \sqrt{{r}^2-R^2}}{R^2 {r}^2}.
\end{equation}

\noindent Using Eq. \eqref{shapiro-time-delay-general_2}, again using the fact that $R_{\odot}/r_{\oplus} << 1$ and $R_{\odot}/r_{\mercury} << 1$, the time delay becomes
\begin{align}
(\Delta t)_{max} &\simeq 4M \left[1 + \ln \left(\frac{4 r_{\oplus} r_{\mercury}}{{R_{\odot}}^2} \right) \right] +\frac{2 \Lambda}{9} \bigg( {r_{\oplus}}^3 \nonumber \\
&+ {r_{\mercury}}^3  \bigg)+ \frac{1792 \pi  \alpha}{3 {R_{\odot}}^3}-256 \alpha  \Bigg[\frac{({r_{\oplus}}+{r_{\mercury}})}{3 {R_{\odot}}^2 {r_{\oplus}} {r_{\mercury}}} \nonumber \\
&+ \left(\frac{1}{{r_{\oplus}}^3}+\frac{1}{{r_{\mercury}}^3}\right) \Bigg].
\label{shapiro-time-delay_n=3_solution}
\end{align}

From an analysis carried onto higher values of $n$ (up to $n = 8$), a series solution depending on the value of $n$ has been noted. For the case $2 \leq n \leq 6$, and assuming $n$ is a natural number, we get 
\begin{align}
&(\Delta t)_{max} \simeq 4M \left[1 + \ln \left(\frac{4 r_{\oplus} r_{\mercury}}{{R_{\odot}}^2} \right) \right] +\frac{2 \Lambda}{9} \bigg( {r_{\oplus}}^3 \nonumber \\
&+ {r_{\mercury}}^3  \bigg) + \alpha \dfrac{\beta^n \pi}{{R_{\odot}}^{2n-3}}+ \alpha \sum\limits_{k=0}^{n-2} \dfrac{\gamma_k^n}{{R_{\odot}}^{2(n-k-2)}} \Bigg(\dfrac{1}{{r_{\oplus}}^{2k+1}} \nonumber \\
&+ \dfrac{1}{{r_{\mercury}}^{2k+1}} \Bigg),
\label{shapiro-time-delay_less-than-6_solution}
\end{align}

\noindent where $\beta^n$ and $\gamma_k^n$ are constants. On the other hand, for $n \geq 7$ and keeping $n$ to be a natural number, an extra contribution appears. Thus, for higher order values of $n$, more contributions to the time delay might appear and a general solution would be difficult to obtain. Nonetheless, the form obtained for the cases between $2 \leq n \leq 6$ are consistent even for $n \geq 7$. In particular, it was noted that for $n \geq 3$, we have $\gamma_0^n = -2^{3n-2}/(2n-3)$. 

Using the solutions found for $n = 2$ and $n = 3$, we can set a constraint on $\alpha$ using the $\gamma$ PPN parameter as follows. As shown in \cite{weinberg1972gravitation}, the maximum time delay in the PPN formalism yields,
\begin{equation}
(\Delta t)_{max}^{PPN} \simeq 4M \left[1 + \dfrac{1+\gamma}{2} \ln \left(\frac{4 r_{\oplus} r_{\mercury}}{{R_{\odot}}^2} \right) \right].
\label{shapiro-time-delay_PPN}
\end{equation}

From the analysis of the Cassini experiment, a value of $\gamma = 1.000021 \pm 0.000023$ was obtained \cite{Bertotti:2003rm}. Taking $\Lambda \sim 10^{-46} \: \text{km}^{-2}$, for the case when $n = 2$, we obtain
\begin{equation}
-9.175 \times 10^{-2} < \alpha < 2.019 \: \text{km}^{2},
\label{time_delay-alpha-contrainst-n2}
\end{equation} 

\noindent whilst for $n = 3$, we obtain
\begin{equation}
-1.183 \times 10^{10} < \alpha < 2.603 \times 10^{11} \: \text{km}^{4}.
\label{time_delay-alpha-contrainst-n3}
\end{equation} 

\section{VI. Gravitational Redshift}\label{sec:grav-redshift}

Consider a stationary clock in a gravitational field. The proper time measured by the clocks is given by the proper time $d\tau$, which from Eq. \eqref{metric-ansatz} becomes,
\begin{equation}
d\tau = e^{A(r)/2} \: dt.
\end{equation}

\noindent To measure the time dilation (and hence the redshift), consider two clocks at rest placed at two positions $r_1$ and $r_2$. The ratio of the frequency (observed at $r_1$) of the light from the point at $r_2$ to that of the light from the point at $r_1$ is
\begin{equation}
\dfrac{\nu_2}{\nu_1} = \dfrac{e^{A(r_2)/2}}{e^{A(r_1)/2}}.
\end{equation}

\noindent Thus, the redshift $z$ is found to be
\begin{equation}
z \equiv \dfrac{\nu_2}{\nu_1} - 1 = \dfrac{e^{A(r_2)/2}}{e^{A(r_1)/2}} - 1.
\end{equation}

\noindent For $|A^{(1)}(r_1)| < 1$ and $|A^{(1)}(r_2)| < 1$, expanding the previous expression up to first order in $M$, $\alpha$ and $\Lambda$, and neglecting their products, the redshift is found to be
\begin{align}
z \simeq \frac{M}{{r_1}}&-\frac{M}{{r_2}} +\frac{\Lambda}{6} \left({r_1}^2-{r_2}^2\right) \nonumber \\
&- \alpha\frac{2^{3 n-2} \left({r_1}^{2-2 n}-{r_2}^{2-2 n}\right)}{3-2 n}. \label{weinberg-redshift-soln}
\end{align}

The first two terms in the expression for the redshift are the Schwarzchild de-Sitter contributions \cite{Kagramanova:2006ax}, whilst the $\alpha$ term is the new contribution which arises from the metric. For the case when $n = 0$ and using the transformation Eq. \eqref{new-cosmological-constant-n0}, the solution reduces to the standard Schwarzchild de-Sitter solution with $k$ as the new cosmological constant, as expected. 

In order to use this result to constrain $\alpha$, we use the experimental setup used by Pound and Rebka which determined the redshift due to the Earth on a particle falling a height of $22.5$ m \cite{Pound:1959}. 

The redshift measured was $z = (2.57 \pm 0.26) \times 10^{-15}$. Using Eq. \eqref{weinberg-redshift-soln} with $\Lambda \sim 10^{-46} \: \text{km}^{-2}$, for $n = 2$ we get
\begin{equation}
-5.052 \times 10^{-5} < \alpha < 1.369 \times 10^{-4} \: \text{km}^{2},
\label{redshift-alpha-contrainst-n2}
\end{equation}

\noindent whilst for the case when $n = 3$, we get
\begin{equation}
-3.854 \times 10^2 < \alpha < 1.044 \times 10^3 \: \text{km}^{4}.
\label{redshift-alpha-contrainst-n3}
\end{equation}

\section{VII. Conclusion}\label{sec:Conclusion}

The equations Eqs. \eqref{perihelion-general-solution}, \eqref{one-sided-deflection_n2}, \eqref{one-sided-deflection_n3}, \eqref{one-sided-deflection_n}, \eqref{one-sided-deflection_ngeq4}, \eqref{shapiro-time-delay_n=2_solution}, \eqref{shapiro-time-delay_n=3_solution}, \eqref{shapiro-time-delay_less-than-6_solution} and \eqref{weinberg-redshift-soln} are the main results of this paper. 

In all tests, the Schwarzchild de-Sitter contributions were obtained, and the effect of $\alpha$ were considered. Although the modification terms were clearly obtained for perihelion precession and gravitational redshift, the light bending and Shapiro time delay are incomplete for general $n$. In the case of light bending, the coefficient of the modification term can only be determined for specific values of $n$ whilst for Shapiro time delay, a power series solution is noted but is incomplete since larger values of $n$ gives rise to contributions not considered in this series solution.

Nonetheless, a set of constraints on $\alpha$ from each test using observational data were obtained for the cases $n = 2$ and $n = 3$, given in Eqs. \eqref{perihelion-alpha-contrainst-n2}, \eqref{perihelion-alpha-contrainst-n3}, \eqref{light_bending-alpha-contrainst-n2}, \eqref{light_bending-alpha-contrainst-n3}, \eqref{time_delay-alpha-contrainst-n2}, \eqref{time_delay-alpha-contrainst-n3}, \eqref{redshift-alpha-contrainst-n2} and \eqref{redshift-alpha-contrainst-n3}. By comparing all the constraints, the values of $\alpha$ for $n=2$ and 3 which satisfy all the solar system tests (in other words, lie within the constraint of each test) are 
\begin{equation}
-5.052 \times 10^{-5} < \alpha < -8.072 \times 10^{-15} \: \text{km}^{2},
\end{equation}

\noindent for $n = 2$ and
\begin{equation}
-3.854 \times 10^2 < \alpha < -25.753 \: \text{km}^{4},
\end{equation} 

\noindent for $n = 3$. Note that in both cases, $\alpha$ is found to be negative.

Although there is a large difference in magnitude between the values of $\alpha$ for the two cases of $n$ considered in this paper, it still keeps the $f(T)$ modification term $\alpha T^n$ small. For example, noting the form of the torsion scalar in Eq. \eqref{torsion-scalar-metric}, for $n = 3$, $T$ is inversely proportional to $r$. Thus, the largest value of $T$ would be obtained by taking the smallest value of the latter. 

In the case of perihelion precession, this corresponds to the semi-latus rectum of Mercury $\mathcal{L}_{\text{M}}$. Using the obtained constraint $|\alpha| \sim 10^{2} \: \text{km}^4$, we get $\big|\alpha {T(\mathcal{L}_{\text{M}})}^3 \big| \sim 10^{-42} \: \text{km}^{-2}$ which is much smaller than $\big|T(\mathcal{L}_{\text{M}}) \big| \sim 10^{-15} \: \text{km}^{-2}$. 

\section*{Acknowledgements}

We would like to thank the unknown referee for the useful comments and suggestions that helped us to significantly improve our manuscript. The research work disclosed in this publication is partially funded by the ENDEAVOUR Scholarships Scheme.

\end{document}